\documentstyle[epsfig,12pt]{article}

\newcommand{\be}{\begin{equation}}
\newcommand{\ee}{\end{equation}}
\newcommand{\ba}{\begin{eqnarray}}
\newcommand{\ea}{\end{eqnarray}}
\newcommand{\bd}{\begin{displaymath}}
\newcommand{\ed}{\end{displaymath}}

\def\twoth{{\textstyle{\frac{2}{3}}}}

\def\fourth{{\textstyle{\frac{4}{3}}}}

\title{
{\bf Strongly Interacting Low Viscosity Matter Created in Heavy Ion Collisions}}
\author{Joseph I. Kapusta\\
{\it School of Physics and Astronomy, University of Minnesota}\\
 {\it Minneapolis, Minnesota 55455, USA}}

\date{\empty}

\begin{document}

\maketitle

\begin{abstract}

Substantial collective flow is observed in collisions between large nuclei at 
high energy, as evidenced by single-particle transverse momentum distributions and by azimuthal correlations among the produced particles.  The data are 
well-reproduced by perfect fluid dynamics.  In a separate development, calculation of the dimensionless ratio of shear viscosity $\eta$ to entropy density $s$ by Kovtun, Son and Starinets within AdS/CFT yields $\eta/s = 1/4\pi$, and they conjectured that this is a lower bound for any physical system.  It is shown that the transition from hadrons to quarks and gluons has behavior similar to helium, nitrogen, and water at and near their phase transitions in the ratio $\eta/s$. Therefore it is possible that experimental measurements can pinpoint the location of this transition or rapid crossover in QCD via the viscosity to entropy ratio in addition to and independently of the equation of state.   
  
\end{abstract}

A transition from a phase of hadrons to a phase of quarks and gluons with increasing temperature and/or baryon chemical potential has been studied theoretically for three decades \cite{KMR}.  Calculations with effective hadronic field theories and with perturbative QCD give consistent and intuitively reasonable estimates of where this transition occurs, but at the present time only lattice gauge theory calculations on finite lattices of finite size can yield quantitatively accurate numbers.  For two flavors of massless quarks the phase transition is of second order.  For three flavors of massless quarks the phase transition is of first order.  For the real world with nonzero masses for up, down and strange quarks the answer is not definitively known yet, but the answer is likely to be a rapid crossover from one phase to another without a rigorous thermodynamic phase transition, at least for zero baryon chemical potential.  Indeed, there may be a line of first order phase transition in the plane of temperature $T$ versus baryon chemical potential $\mu$ starting from the chemical potential axis and terminating at some critical point in the $T - \mu$ plane \cite{Forcrand}.  The Relativistic Heavy Ion Collider (RHIC) at Brookhaven National Laboratory was constructed explicitly to create quark gluon plasma.  After more than five years of operation, what have the experiments told us?    

One of the amazing discoveries of experimental measurements of gold on gold 
collisions at RHIC is the surprising amount of collective flow exhibited by the outgoing hadrons.  Collective flow is evidenced in both the single-particle transverse momentum distribution \cite{radial}, commonly referred to as radial flow, and in the asymmetric azimuthal distribution around the beam axis \cite{RHICv2}, quantified by the functions $v_1(y,p_T), v_2(y,p_T), ...$ in the expansion
\be
\frac{d^3N}{dydp_Td\phi} = \frac{1}{2\pi}\frac{d^2N}{dydp_T} \left[ 1 
+ 2v_1(y,p_T) \cos(\phi) + 2v_2(y,p_T) \cos(2\phi) + \cdot\cdot\cdot \right]
\ee
where $y$ is the rapidity and $p_T$ is the transverse momentum.  The function 
$v_2(y=0,p_T)$, in particular, was expected to be much smaller at RHIC than it 
was at the lower energies of the SPS (Super Proton Synchrotron) at CERN 
\cite{SPSv2}; in fact, it is about twice as large.  Various theoretical calculations \cite{coalesce} support the notion that collective flow is mostly generated early in the nucleus-nucleus collision and is present before partons coalesce or fragment into hadrons.  Theoretical calculations including only 
two-body interactions between partons cannot generate sufficient flow to explain the observations unless partonic cross sections are artificially enhanced by more than an order of magnitude over perturbative QCD predictions \cite{Molnar}.  Thus quark-gluon matter created in these collisions is strongly interacting, unlike the type of weakly interacting quark-gluon plasma expected to occur at very high temperatures on the basis of asymptotic freedom \cite{pQCD}.  On the other hand, lattice QCD calculations yield an equation of state that differs from an ideal gas only by about 10\% once the temperature exceeds $1.5 T_c$, where $T_c \approx 175$ MeV is the critical or crossover temperature from quarks and gluons to hadrons \cite{latticeQCD}.  Furthermore, perfect fluid dynamics with zero shear and bulk viscosities reproduces the measurements of radial flow and $v_2$ very well up to transverse momenta of order 1.5 GeV/c \cite{hydro}.  Parametric fits to the transverse momentum spectra of hadrons, such as pions, kaons, and protons, result in average transverse fluid flow velocities of more than 50/

An amazing theoretical discovery was made by Kovtun, Son and Starinets \cite{Son}.  They showed that certain special field theories, special in the sense that they are dual to black branes in higher space-time dimensions, have the ratio $\eta/s = 1/4\pi$ (in units with $\hbar = k_{\rm B} = c =1$) where $\eta$ is the shear viscosity and $s$ is the entropy density.  The shear viscosity is rigorously given by the Kubo formula
\be
\eta = \frac{1}{20} \lim_{\omega \rightarrow 0}
\frac{1}{\omega} \int d^4x e^{i\omega t} \langle
\left[ T^{ij}_{\rm traceless}(x), T^{ij}_{\rm traceless}(0) \right] \rangle
\ee
The connection between transport coefficients and gravity is intuitively clear since both involve (commutators of) the stress-energy-momentum tensor $T^{ij}$.
They conjectured that all substances have this value as a lower limit, and gave as examples helium, nitrogen, and water at pressures of 0.1 MPa, 10 MPa, and 100 MPa, respectively.  Is the RHIC data, represented by radial and elliptic flow, telling us that the created matter has a very small viscosity, that it is a {\it perfect fluid}?

The relatively good agreement between perfect fluid calculations and experimental data for hadrons of low to medium transverse momentum at RHIC suggests that the viscosity is small; however, it cannot be zero.  Indeed, the calculations within AdS/CFT suggests that $\eta \ge s/(4\pi)$.  It will now be argued that sufficiently precise calculations and measurements should allow for a determination of the ratio $\eta/s$ as a function of temperature, and that this ratio can pinpoint the location of the phase transition or rapid crossover from hadronic to quark and gluon matter.  This is a different method than trying to infer the equation of state of QCD in the form of pressure $P$ as a function of temperature $T$ or energy density $\epsilon$.

What happens in a gas as compared to a liquid?  The simplest way to understand the general behavior was presented by Enskog, as summarized in \cite{atomtheory}.  Shear viscosity represents the ability to transport momentum.  In the classical transport theory of gases $\eta/s \sim T l_{\rm free} \bar{v}$, where $l_{\rm free}$ is the mean free path and $\bar{v}$ is the mean speed.  For a dilute gas the mean free path is large, $l_{\rm free} \sim 1/n\sigma$, with $n$ the particle number density and $\sigma$ the cross section.  Hence it is easy for a particle to carry momentum over great 
distances, leading to a large viscosity.  (This is the usual paradox, that a nearly ideal classical gas has a divergent viscosity.)  In a liquid there are strong correlations between neighboring atoms or molecules.  A liquid is homogeneous on a mesoscopic scale, but on a microscopic scale it is a mixture of clusters and voids.  The action of pushing on one atom is translated to the next one and so on until a whole row of atoms moves to fill a void, thereby transporting momentum over a relatively large distance and producing a large viscosity.  Reducing the temperature at fixed pressure reduces the density of voids, thereby increasing the viscosity.  (This is commonly experienced with motor oil in the cold winter months.)  The viscosity, normalized to the entropy, is expected to be the smallest at or near a critical temperature, corresponding to the most difficult condition to transport momentum.  What do the atomic and molecular data show? 

In figures 1 and 2 are plotted the ratio $\eta/s$ versus temperature at three 
fixed pressures, one of them being the critical pressure (meaning that the curve passes through the critical point) and the other ones being larger and smaller, for helium and water.  (Other substances, such as nitrogen, behave similarly.)  The ratios were constructed with data obtained from the National Institute of Standards and Technology (NIST) \cite{NIST}.  (Care must be taken to absolutely normalize the entropy to zero at zero temperature; we did that using data from CODATA \cite{CODATA}.)  The important observation \cite{private,singular} is that $\eta/s$ has a minimum at the critical point where there is a cusp.  At pressures below the critical pressure there is a discontinuity in $\eta/s$, and at pressures above it there is a broad smooth minimum.  This is an empirical observation.  Figure 3 shows a plot of $\eta/s$ (in units of its minimum value at the critical point) versus $T/T_c$ for helium, nitrogen, and water along their critical isobars.  Unfortunately there is no obvious universal scaling law.

Numerical many-body simulations have been performed for a variety of systems.  The result of a calculation for a 2-dimensional Yukawa system in the liquid state is shown in figure 4 \cite{Goree}.  Here $\Gamma = Q^2/aT$ is the Coulomb coupling parameter with $Q$ the electric charge and $a$ the Wigner-Seitz radius.  It displays explicitly the dominance of potential contributions at low temperature and the dominance of kinetic contributions at high temperature, in agreement with the intuition provided by Enskog.

The energy-momentum tensor density for a perfect fluid (which does not imply 
that the matter is non-interacting) is $T^{\mu\nu} = -
Pg^{\mu\nu}+wu^{\mu}u^{\nu}$.  Here $w=P+\epsilon=Ts$ is the local enthalpy density and $u^{\mu}$ is the local flow velocity.  Corrections to this expression are proportional to first derivatives of the local quantities whose coefficients are the shear viscosity $\eta$ and bulk viscosity $\zeta$.  (Thermal conductivity is neither relevant nor defined when all net conserved charges, such as electric charge and baryon number, are zero.)  Explicit expressions are to be found in textbooks \cite{fluid} and they are quite lengthy.  There is an ambiguity to the meaning of flow velocity for relativistic dissipative systems.  In the Eckart approach $u^{\mu}$ is the 4-velocity for baryon transport.  In the Landau-Lifshitz approach it is the 4-velocity for energy transport.  Neither approach is more correct than the other.  Both yield the same divergence of the entropy density current.
\be
\partial_{\mu} s^{\mu} = \frac{\eta}{2T} \left( \partial_i u^j + \partial_j u^i
- \twoth \delta_{ij} \partial_k u^k \right)^2
+ \frac{\zeta}{T} \left( \partial_k u^k \right)^2
+ \frac{\chi}{T^2} \left( \partial_k T + T \dot{u}_k \right)^2
\ee
Here $\chi$ is the coefficient of thermal conductivity.  This expression embodies the non-decrease of entropy.

Perfect fluid dynamics applies when the viscosities are small, or when the gradients are small, or both.  The dispersion relations for the transverse and longitudinal (pressure) parts of the momentum density are
\ba
\omega + i D_t k^2 &=& 0 \nonumber \\
\omega^2 - v^2 k^2 + i D_l \omega k^2 &=& 0
\ea
where $D_t = \eta/w$ and $D_l = (\fourth \eta + \zeta)/w$ are diffusion 
constants with the dimension of length and $v$ is the speed of sound.  Since 
$w=Ts$, and since usually the bulk viscosity is small compared to the shear 
viscosity, the dimensionless ratio of (shear) viscosity to entropy (disorder) 
$\eta/s$ is a good way to characterize the intrinsic ability of a substance to 
relax towards equilibrium independent of the actual physical conditions 
(gradients of pressure, energy density, etc.).  It is also a good way to compare 
very different substances.

How does this relate to hadrons and quark-gluon plasma?  In the low energy 
chiral limit for pions the cross section is proportional to $\hat{s}/f_{\pi}^4$, 
where $\hat{s}$ is the usual Mandelstam variable for invariant mass-squared and 
$f_{\pi}$ is the pion decay constant.  The thermally averaged cross section is 
$\langle \sigma \rangle \propto T^2/f_{\pi}^4$, which leads to $\eta/s \propto 
(f_{\pi}/T)^4$.  Explicit calculation gives \cite{Prakash}
\be
\frac{\eta}{s} = \frac{15}{16\pi} \frac{f_{\pi}^4}{T^4}
\label{chiral}
\ee
Thus the ratio $\eta/s$ diverges as $T \rightarrow 0$.  At the other extreme 
lies quark-gluon plasma.  The parton cross section behaves as $\sigma \propto 
g^4/\hat{s}$.  A first estimate yields $\eta/s \propto 1/g^4$.  Asymptotic 
freedom at one loop order gives $g^2 \propto 1/\ln(T/\Lambda_T)$ where 
$\Lambda_T$ is proportional to the scale parameter $\Lambda_{\rm QCD}$ of QCD.  
Therefore $\eta/s$ is an increasing function of $T$ in the quark-gluon phase.  
As a consequence, $\eta/s$ must have a minimum.  Based on atomic and molecular 
data, this minimum should lie at the critical temperature if there is one, 
otherwise at or near the rapid crossover temperature.

The most accurate and detailed calculation of the viscosity in the low 
temperature hadron phase was performed in \cite{Prakash}.  The two-body 
interactions used went beyond the chiral approximation, and included 
intermediate resonances such as the $\rho$-meson.  The results are displayed in 
figure 5, both two flavors (no kaons) and three flavors (with kaons).  The 
qualitative behavior is the same as in eq. (\ref{chiral}).  The most accurate 
and detailed calculation of the viscosity in the high temperature quark-gluon 
phase was performed in \cite{Arnold}.  They used perturbative QCD to calculate 
the full leading-order expression, including summation of the Coulomb 
logarithms.  For three flavors of massless quarks the result is
\be
\frac{\eta}{s} = \frac{5.12}{g^4 \ln(2.42/g)}
\ee
We used this together with the two-loop renormalization group expression for the 
running coupling
\be
\frac{1}{g^2(T)} = \frac{9}{8\pi^2} \ln\left( \frac{T}{\Lambda_T} \right)
+ \frac{4}{9\pi^2} \ln \left( 2 \ln\left( \frac{T}{\Lambda_T} \right) \right)
\ee
with $\Lambda_T = 30$ MeV, which approximately corresponds to using an energy 
scale of $2\pi T$ and $\Lambda_{\overline{MS}} = 200$ MeV.  The result is also 
plotted in figure 5.  These results imply a minimum in the neighborhood of the 
expected value of $T_c \approx 190$ MeV.  Whether there is a discontinuity or a 
smooth crossover cannot be decided since both calculations are unreliable near 
$T_c$.

It is interesting to ask what happens in the large $N_c$ limit with $g^2N_c$ 
held fixed \cite{largeN}.  In this limit, meson masses do not change very much 
but baryon masses scale proportional to $N_c$; therefore, baryons may be 
neglected in comparison to mesons due to the Boltzmann factor.  Since the meson 
spectrum is essentially unchanged with increasing $N_c$, so is the Hagedorn 
temperature.  The critical temperature to go from hadrons to quarks and gluons 
is very close to the Hagedorn temperature, so that $T_c$ is not expected to 
change very much either.  In the large $N_c$ limit the meson-meson cross section 
scales as $1/N_c^2$.  According to our earlier discussion on the classical 
theory of gases, this implies that the ratio $\eta/s$ in the hadronic phase 
scales as $N_c^2$.  This general result is obeyed by (\ref{chiral}) since it is 
known that $f_{\pi}^2$ scales as $N_c$.  The large $N_c$ limit of the viscosity 
in the quark and gluon phase may be inferred from the calculations of 
\cite{Arnold} to be
\be
\left(\frac{\eta}{s}\right)_{\rm QGP} = 
\frac{A}{\left(g^2N_c\right)^2 \ln \left( B/(g^2N_c)\right)}
\ee
where $A$ and $B$ are known constants but dependent upon $N_f/N_c$.  Thus the ratio $\eta/s$ has a finite large $N_c$ limit in 
the quark and gluon phase.  Therefore, we conclude that $\eta/s$ has a 
discontinuity proportional to $N_c^2$ if $N_c \rightarrow \infty$.  This jump is 
in the opposite direction to that in figure 5.

So far the only quantitative results for viscosity in lattice gauge theory have 
been reported by Nakamura and Sakai \cite{latticeeta} for pure SU(3) without 
quarks.  This bold effort obtained $\eta/s \approx 1/2$ in the temperature range 
$1.6 < T/T_c < 2.2$, albeit with uncertainties of order 100\%.  Gelman, Shuryak 
and Zahed \cite{cQCD} have modeled the dynamics of long wavelength modes of QCD 
at temperatures from $T_c$ to $1.5 T_c$ as a classical, nonrelativistic gas of 
massive quasi-particles with color charges.  They obtained a ratio of $\eta/s 
\approx 0.34$ in this temperature range.

Another interesting approach has been the computation of $\eta/s$ in 
${\cal N}=4$ supersymmetric $SU(N_c)$ Yang-Mills theory (SYM).  In that theory it is possible to do calculations in the large coupling limit \cite{SUSY} and in the weak coupling limit \cite{Huot}, and without too much imagination it is possible to find a paramtrization that interpolates smoothly between the two limits.  However, SYM has no renormalization group running coupling, no asymptotic freedom, and no thermodynamic phase transition.  Since it has so many more degrees of freedom than QCD as possible scattering targets, its viscosity to entropy ratio is much smaller than QCD at high temperature when compared at the same value of the gauge coupling.  However, when the theories are compared at the same value of the Debye screening mass they do agree reasonably well \cite{Huot}.

It ought to be possible to extract numerical values of the viscosity in heavy ion collisions via scaling violations to perfect fluid flow predictions.  The program is to solve relativistic viscous fluid equations, with appropriate initial conditions and with a hadron cascade afterburner, over a range of beam energies and nuclei and extract $\eta(T)/s(T)$ from comparison with data.  This program is analogous to what was accomplished at lower energies of 30 to 1000 MeV per nucleon beam energies in the lab frame.  At those energies, it was possible to infer the compressibility of nuclear matter and the
momentum-dependence of the nuclear optical potential via the transverse momentum distribution relative to the reaction plane \cite{BUU} and via the balance between attractive and repulsive scattering \cite{balance}.

At RHIC some of the specific proposals to extract or infer the viscosity to entropy ratio from data include: elliptic flow, Hanbury-Brown and Twiss (HBT) interferometry, single particle momentum spectra, and momentum fluctuations.  Other possibilities which have not been worked on yet include jet quenching and photon and dilepton spectra.  Some of the complications include the possibility that gradients are so large that the second-order dissipative equations of Israel and Stewart are necessary and that turbulence in the plasma may lead to an anomalous viscosity.  Clearly this is an interesting and challenging goal but worth the effort!

This work was supported by the US Department of Energy under grant DE-FG02-87ER40328.

\newpage

\begin{figure}
 \centering
 \includegraphics[width=3.5in,angle=90]{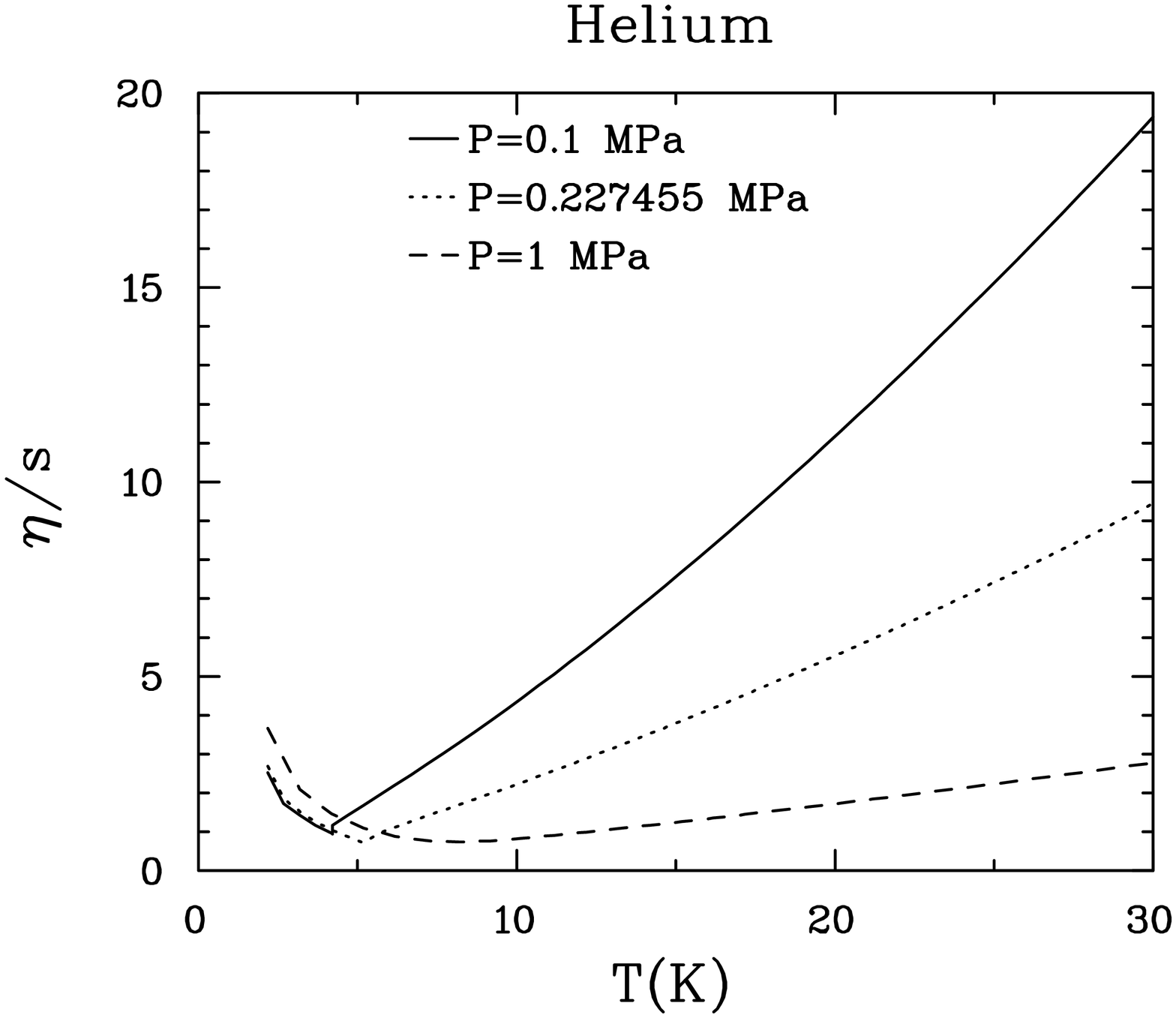}
 \caption{The ratio $\eta/s$ as a function of $T$ for helium with $s$ normalized such that $s(T=0)=0$.  The curves correspond to fixed pressures, one of them being the critical pressure, and the others being greater (1 MPa) and the other smaller (0.1 MPa).  Below the critical pressure there is a jump in the ratio, and above the critical pressure there is only a broad minimum. They were constructed using data from NIST and CODATA.}
\end{figure}

\begin{figure}
 \centering
 \includegraphics[width=3.5in,angle=90]{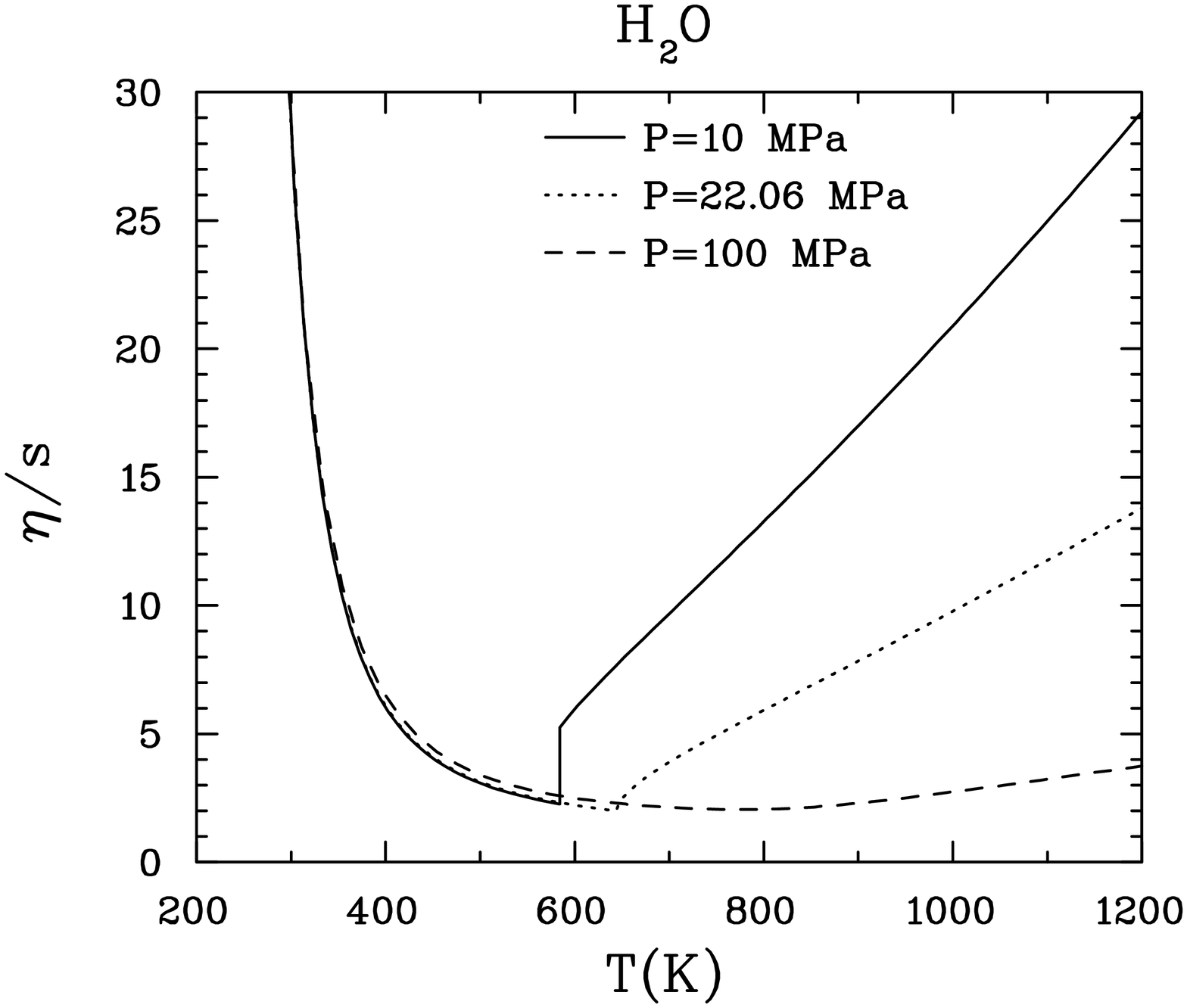}
 \caption{ The ratio $\eta/s$ as a function of $T$ for water with $s$ normalized such that $s(T=0)=0$.  The curves correspond to fixed pressures, one of them being the critical pressure, and the others being greater (100 MPa) and the other smaller (10 MPa).  Below the critical pressure there is a jump in the ratio, and above the critical pressure there is only a broad minimum. They were constructed using data from NIST and CODATA.}
\end{figure}

\begin{figure}
 \centering
 \includegraphics[width=3.5in,angle=90]{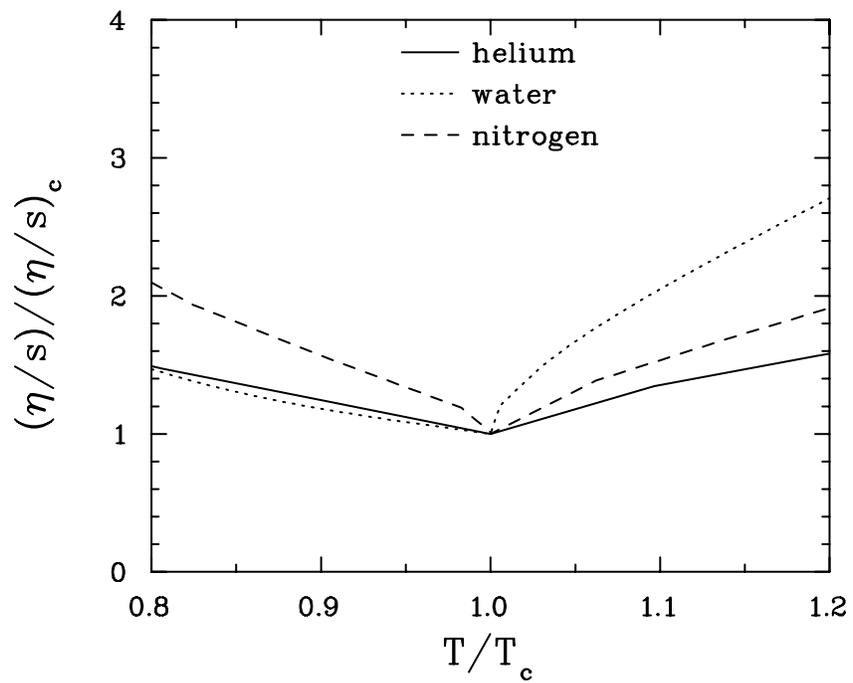}
 \caption{ The ratio $\eta/s$ in units of its value at the critical point as a function of $T/T_c$ for three common systems.  The curves are isobars at the critical pressure.  They were constructed using data from NIST and CODATA.}
\end{figure}

\begin{figure}
 \centering
 \includegraphics[width=3.5in,angle=0]{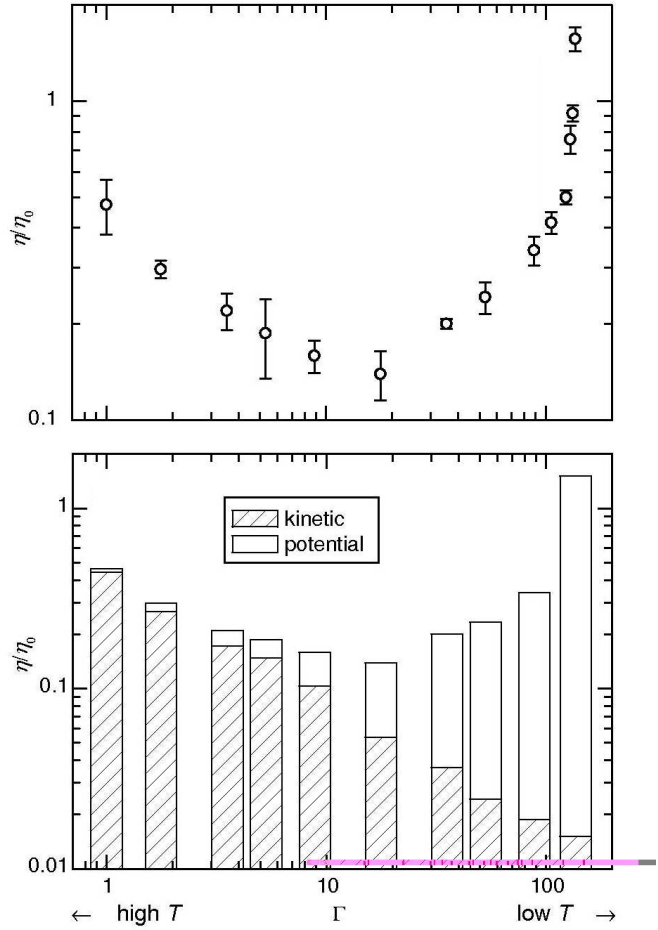}
 \caption{ The shear viscosity versus Coulomb parameter for a two dimensional Yukawa liquid (dusty plasma) from simulations by Liu and Goree \cite{Goree}.  Potential energy dominates at low temperatures and kinetic energy dominates at high temperatures.}
\end{figure}

\begin{figure}
 \centering
 \includegraphics[width=3.5in,angle=90]{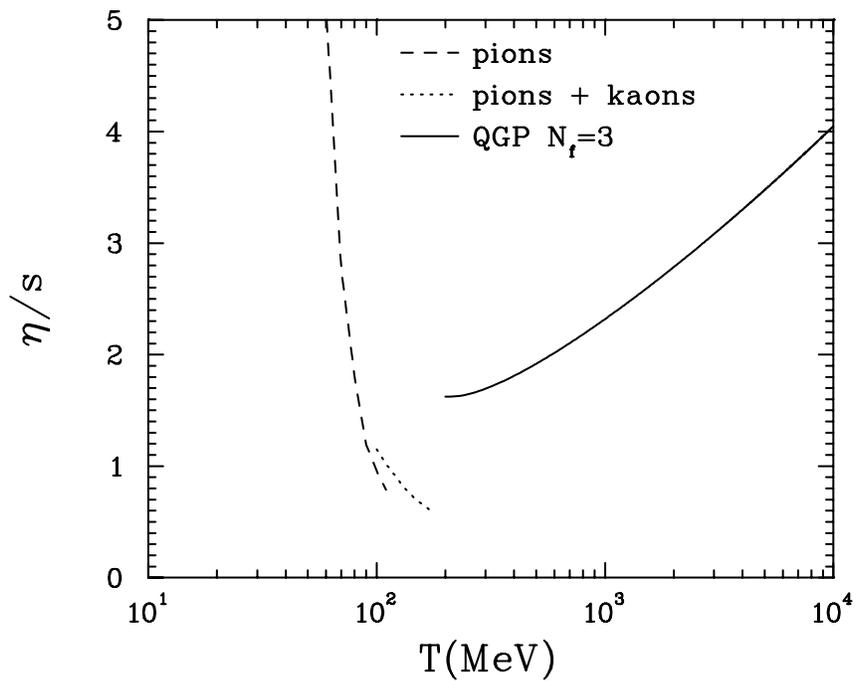}
 \caption{The ratio $\eta/s$ for the low temperature hadronic phase and for the 
high temperature quark-gluon phase.  Neither calculation is very reliable in the 
vicinity of the critical or rapid crossover temperature.}
\end{figure}

\end{document}